\documentclass[aps,prd,twocolumn,nofootinbib,longbibliography,10pt]{revtex4}
\usepackage{etoolbox}
\usepackage{dcolumn,tensor,nicefrac}
\usepackage{amsmath,amssymb,amsfonts,mathtools}
\usepackage{mathrsfs,bbold}
\usepackage{graphicx}
\usepackage[colorlinks=true,urlcolor=blue,citecolor=red,linkcolor=blue]{hyperref}
\usepackage{accents}
\newlength{\dhatheight}
\usepackage{natbib}
\usepackage{wrapfig}
\usepackage{booktabs}
\usepackage{float}
\usepackage{flushend,BOONDOX-cal,BOONDOX-frak}
\newcommand{\paren}[1]{\left(#1\right)}
\newcommand{\second}[1]{\left\{#1\right\}}
\newcommand{\third}[1]{\left[#1\right]}

\makeatletter
\newsavebox{\@brx}
\newcommand{\llangle}[1][]{\savebox{\@brx}{\(\m@th{#1\langle}\)}%
  \mathopen{\copy\@brx\kern-0.5\wd\@brx\usebox{\@brx}}}
\newcommand{\rrangle}[1][]{\savebox{\@brx}{\(\m@th{#1\rangle}\)}%
  \mathclose{\copy\@brx\kern-0.5\wd\@brx\usebox{\@brx}}}
\makeatother
\begin{document}
\title{\textbf{Quantum corrections to the black hole entropy using the brick wall model}}
\author{Gopinath Guin}
\email{ gopinath.guin@bose.res.in}
\affiliation{Department of Astrophysics and High Energy Physics, S. N. Bose National Centre for Basic Sciences, JD Block, Sector-III, Salt Lake City, Kolkata-700 106, India}
\author{Arpita Jana}
\email{janaarpita2001@gmail.com}
\thanks{Gopinath Guin and Arpita Jana contributed equally to this work.}
\affiliation{Department of Astrophysics and High Energy Physics, S. N. Bose National Centre for Basic Sciences, JD Block, Sector-III, Salt Lake City, Kolkata-700 106, India}

\author{Sunandan Gangopadhyay}
\email{sunandan.gangopadhyay@gmail.com}
\affiliation{Department of Astrophysics and High Energy Physics, S. N. Bose National Centre for Basic Sciences, JD Block, Sector-III, Salt Lake City, Kolkata-700 106, India}

\begin{abstract}
\noindent It was suggested by 't Hooft that to avoid the divergence of the free energy or entropy of the probe fields, one has to introduce an infrared (IR) cutoff. This introduction of the IR cutoff, namely, the brick wall model, has been studied for the Schwarzschild black hole and has been shown by 't Hooft that to have the entropy of the order of Bekenstein-Hawking entropy, the cutoff must be of the order of Planck length. In this study, we investigated the brick wall model  in the context of a Reissner–Nordstr\"om, quantum corrected Schwarzschild and quantum corrected Reissner-Nordstr\"om  black holes. In particular, we have shown that the brick wall model can give logarithmic correction to the black hole entropy in the subleading order. The important aspect of our analysis is that we have not considered the near horizon approximation of the lapse function and get more accurate correction to the entropy. {This corrected entropy has an important feature that it is larger than the usual Bekenstein-Hawking entropy which aligns with the result of Barrow entropy.}
\end{abstract}
\maketitle
\section{introduction}
\noindent The intensive study of thermodynamic properties and the entropy of black hole has been a rich domain of exploration in black hole Physics nowadays. Bekenstein, in his classic works \cite{Bekenstein1, Bekenstein2}, showed that the entropy associated to the black hole is proportional to its horizon area, giving the firm footing to Hawking's pioneering discovery that the black holes radiate thermally with a certain temperature $T_H$, later named as the Hawking temperature \cite{Hawking1, Hawking2,Hawking3}. Till now, several methods have been developed to calculate the black hole entropy. In string theoretic approach, the entropy for extremal and near-extremal black holes have been calculated by counting the microstates for D-branes \cite{strominger,johnsonkhuri,maldacena,banerjeegupta1,banerjeegupta2,callan,horowitz}. Apart from these, the black hole entropy and its quantum corrections have been derived in different approaches like Euclidean method \cite{gibbons,york,braden,alwis1,alwis2,sen}, conformal anomaly methods \cite{carlip1,carlip2,caicao} and quantum tunneling approach \cite{srinivasan,banerjeemajhi}. There is another commonly used method to compute black hole entropy is the \textit{brick wall method}, first introduced by Gerard 't Hooft in 1984  \cite{brickwall}.\\
\noindent In this article, we have focused on the brick wall formalism, a semi-classical approach, where the gravitational field is treated classically and the other fields propagating outside the event horizon are tackled quantum mechanically. The density of states corresponding to these quantum fields diverge in the vicinity of the black hole, which is the reason for the divergent statistical entropy just outside the event horizon \cite{susskind,mann,demers,barbon}. In the context of brick wall formalism, 't Hooft discussed that this divergence can be regulated introducing a boundary (brick wall) just outside the event horizon of the black hole. Initially, this approach was done for Schwarzschild geometry. The basic assumption for this approach was that the wave functions must all vanish within the fixed distance $h$ (height of the brick wall) from the event horizon,
\begin{equation}
    \varphi(r)=0\, ,\forall ~~r\leq 2M+h\, ,
\end{equation}
where $M$ is the mass of the black hole and $\varphi (r)$ is the scalar wave function for a lightweight spinless particle. Another important condition in this formalism is to introduce IR cutoff in the form of a large box of length $L$, where the field mode satisfies the boundary condition
\begin{equation}
    \varphi(r)=0~, ~~~if~~~ r=L.
\end{equation}
The statistical entropy obtained using this prescription agrees with the Bekenstein-Hawking area law. Now, it is very natural to seek for quantum corrections for entropy using the brick wall formalism and these analyses have been done in \cite{kimkulkarni, winstanley}. In these works, they used near horizon approximation of the lapse function to calculate the free energy, as well as the entropy.\\
\noindent In our work, we seek to compute the corrected entropy in brick wall approach, without using any approximated form of the lapse function and with these, we obtained subleading corrections to the brick wall entropy and got the positive sign in front of the coefficient of the logarithmic term, which makes the quantum corrected entropy larger than the Bekenstein-Hawking entropy.  We have done our analyses for non-extremal black holes only. The application of brick wall model for extremal black holes has been the topic for some controversial studies \cite{alwis2,kimkim,ghoshmitra1,caizhang,ghoshmitra2}. \\
\noindent This article is arranged as following. In section (\ref{brickwall}), we have revisited the theoretical framework for the brick wall formalism and reestablished the form of free energy in this approach, which will lead us to the brick wall entropy. In section (\ref{schwarzschild}), we have computed the entropy correction for the schwarzschild black hole using the brick wall cutoff. While computing the free energy, we have used the exact form of the lapse function, and found the corrected entropy in terms of the height of the brick wall $h$. The leading term of this entropy allows us to equate this with the usual Bekenstein-Hawking entropy, which led us to find the height of the brick wall from the event horizon. Substituting the $h$ back to the entropy expression, we have obtained the final entropy with leading order logarithmic and sub-leading order inverse area corrections along with finite numerical coefficients, independent of any universal constants. In the later sections, we have done the same analysis for Reissner-Nordstr\``om and quantum corrected black hole geometries. In section (\ref{conclusion}), we have presented the important outcomes of our analysis.

\section{theoretical basis of the brick wall model}\label{brickwall}
\noindent The idea of the brickwall model in black hole physics lies in the fact that the number of energy levels any particle can occupy, diverges at the event horizon of the black hole. This divergence occurs because the lapse function $f(r)$ vanishes at the horizon. To avoid this divergence, 't Hooft in \cite{brickwall} introduced an approach in which an IR cutoff was imposed, preserving the basic thermodynamic properties of the black hole.\\
The basic consideration of the brick wall model is that the particle wave function vanishes below the brick wall radius, that is, 
\begin{equation}
    \varphi(r)=0\, ,\forall r\leq r_h+h\, ,
\end{equation}
where $r_h$ is the horizon of the black hole, and $h$ is the thickness of the brick wall. Also another condition is the vanishing of the wave function at the infrared cutoff of the order of a large box length $L$, that is, 
\begin{equation}
    \varphi(L)=0~.
\end{equation}
The Klein-Gordon(KG) equation gives,
\begin{equation}
    f^{-1}(r)E^2\varphi+\frac{1}{r^2}\partial_r\paren{r^2f(r)\partial_r\varphi}-\paren{\frac{l(l+1)}{r^2}+m^2}\varphi=0~.
\end{equation}
Here, $E$ denotes the energy of a state characterized by the quantum numbers $n$, $l$, and $l_z$, where $n$ is the radial quantum number, $l$ is the total angular momentum quantum number, and $l_z$ is the $z$-component of the angular momentum. To proceed further, we can follow the WKB approximation as long as the mass of the black hole $M$ is much larger than the Planck mass\cite{brickwall} to have a radial wave number $\mathcal{k}(E,l,r)$. That gives 
\begin{equation}
    \mathcal{k}^2(E,l,r)=\frac{1}{f(r)}\third{\frac{E^2}{f(r)}-m^2-\frac{l(l+1)}{r^2}} \, .
\end{equation}
This is valid for the classically allowed region $\biggr(E^2 \gg f(r)\paren{m^2+\frac{l(l+1)}{r^2}}\biggr)$, otherwise $\mathcal{k}=0$. The number of radial modes $n$ is given by the quantization condition 
\begin{equation}\label{radial mode}
    \pi n(E,l)=\int_{r_h+h}^Ldr\,  \mathcal{k}(E,r,l)\, .
\end{equation}
Again since each $l$ has $2l+1$ degeneracy, the total number of modes $N$ that lies below the energy $E$ is given by
\begin{align}\label{total number of modes}
    N(E)&=\int dl \, (2l+1) \, n(E,l) \nonumber \\
     &=\frac{1}{\pi}\int dl (2l+1) \int_{r_h+h}^L dr \, \mathcal{k}(E,r,l) \nonumber \\
     &\simeq\frac{2}{3\pi}E^3\int dr \frac{r^2}{f^2(r)}~.
\end{align}
\\
The free energy $F$ of a quantum scalar field in terms of its inverse temperature $\beta$ is given by 
\begin{equation}
    e^{-\beta\,F}=\prod_{n,l,l_z}\frac{1}{1-e^{-\beta\, E}} \, . 
\end{equation}
Inverting this we can write the free energy in terms of the energy $E$ as
\begin{equation}
    F=\frac{1}{\beta}\sum_{n,l,l_z}\ln\paren{1-e^{-\beta\,E}} \, .
\end{equation}
In the continuum limit this reads
\begin{align}
    F&=\frac{1}{\beta}\int dl  \paren{2l+1} \int dn \, \ln\paren{1-e^{-\beta\,E}} \nonumber \\
    &=-\frac{1}{\beta}\int dl  \paren{2l+1} \int dE \, \frac{n(E)\beta}{e^{\beta E}-1}\, ~.
\end{align}
While writing the second step, the boundary terms have not been written as at the ground state $n(E)\rightarrow0$ and at very high energy $e^{-\beta E}\rightarrow0$. Using the expression for $n(E)$ from eq.(s)(\eqref{radial mode},\eqref{total number of modes}), we can write
\begin{align}\label{free energy}
    F&=-\int \frac{dE}{e^{\beta E}-1}N(E)\nonumber\\
    &
    =-\frac{1}{\pi}\int dl \,(2l+1)\int \frac{dE}{e^{\beta E}-1}\int_{r_h+h}^L dr K(E,r,l)\nonumber \\ 
    &\simeq \frac{2}{3\pi}\int dE \frac{E^3}{e^{\beta E}-1}\int_{r_h+h}^L dr \frac{r^2}{f^2(r)} \nonumber \\
   & =-\frac{2\pi^3}{45 \beta ^4}\int_{r_h+h}^L dr \frac{r^2}{f^2(r)}~.
\end{align}
We are now ready to proceed with different types of black hole, by putting its lapse function into the above integral to find out the free energy and then its entropy by the thermodynamic formula
\begin{equation}\label{entropy}
    S_{BW}=\left.\beta^2\frac{\partial F}{\partial \beta}\right|_{\beta=\beta_H}
\end{equation}
where $\beta_H$ is the inverse temperature of the black hole at the horizon. Comparing the entropy result with the known Bekenstein-Hawking area law, we can find the thickness of the brick wall $h$. \\
From our calculation, we have also shown that from the brick wall model, it is possible to derive an expression of the black hole entropy with the ``\textit{universal}" logarithmic correction in subleading order \cite{Kaul,DasMajumdar,gup1,gup2,gup3}.


\section{Brick Wall model for Schwarzschild black hole }\label{schwarzschild}
\noindent In general relativity (GR), the Schwarzschild geometry is an unique metric  describing a spherically symmetric vacuum solution of the Einstein equation and is given by(in the calculation we have used natural units that is $G=c=\hbar=k_B=1$)
\begin{equation}
ds^2 = -f(r) dt^2 + f(r)^{-1} dr^2 + r^2 d\Omega_2^2
\end{equation}
where $f(r)$ is the lapse function and has the form $f(r)=1-\frac{2M}{r}$.
The surface gravity $\kappa$ for a static diagonal metric reads as
\begin{equation}\label{surface gravity schwarzschild}
    \kappa=\left.\frac{1}{2}f'(r)\right|_{r={r_H}}=\frac{1}{4M} \ .
\end{equation}
On the other hand, the surface temperature is related to the surface gravity which is defined as
\begin{equation}\label{horizon temperature}
    T_H=\frac{\kappa}{2\pi}=\frac{1}{8\pi M}~.
\end{equation}
Now, we can write the expression for the free energy using eq.(\ref{free energy}) as
\begin{align}\label{free energy schwarzschild}
F&=\left.-\frac{2\pi^3}{45\beta^4}\int_{r_H+h}^L dr\frac{r^2}{f(r)^2}\right|_{L \to \infty}\nonumber \\&=-\frac{2\pi^3}{45\beta^4}\int_h^\infty dx\frac{(r_{H}+x)^4}{x^2}~.
\end{align}
\noindent Here, in the last line, we have substituted $r=r_{H}+x$.\footnote{In natural units, for Schwarzschild black hole, $r_{H}=2M$.} Performing the integral, we get the expression of free energy as
\begin{align}
    F=-\frac{2\pi^3}{45\beta^4}\biggr[\frac{r_{H}^4}{h}-\frac{h^3}{3}-6r_{H}^2 h-2h^2r_{H}-4r_{H}^3 \ln{h} \biggr]~.
\end{align}
Clearly, one can see the free energy scales as the fourth power of the surface temperature. Now it is very easy to calculate the entropy from the thermodynamics as
\begin{align}\label{entropyBW}
S_{BW}&=\left.\left(\beta^2\frac{\partial F}{\partial\beta}\right)\right|_{\beta=\beta_H}\nonumber\\
&=\frac{1}{2880M^3}\biggr[\frac{(2M)^4}{h}-\frac{h^3}{3}-6(2M)^2 h-4h^2 M \nonumber \\ & ~~~~~~~~~~~~~~~~~-4(2M)^3 \ln{h} \biggr]~.
\end{align}
\noindent Here we have used $\beta_H = T_{H}^{-1}=8\pi M$.
The leading term of this entropy expression must be equal to the Bekenstein-Hawking entropy. So equating this with the Bekenstein-Hawking entropy, $S_{BH}=\frac{A_{BH}}{4l_p^2}$, we get
\begin{align}
& S_{BW}=S_{BH}\nonumber\\
\implies &\frac{M}{180h}={4\pi M^2}\nonumber\\
    \implies & h=\frac{l_p^2}{720\pi M}  =\frac{l_p^2}{360\pi r_H} ~    .
\end{align}
Thus, we obtain an expression for the width of the brick wall, which was first shown in \cite{brickwall}. One can see The width is very small and comparable to the Planck length.\\
\noindent Now substituting this $h$ in eq.(\ref{entropyBW}), we get the corrected brick wall entropy in terms of the horizon area $A_{BH}$ as
\begin{align}
    S_{BW}=&\frac{A_{BH}}{4l_P^2}+\frac{1}{180}\ln{\left( \frac{A_{BH}}{4l_P^2}\right)}-\frac{1}{(30\times 180)}\left(\frac{l_P^2}{A_{BH}} \right)\nonumber\\&-\frac{1}{45\times(180)^2}\left(\frac{l_P^4}{A_{BH}^2} \right)~.
\end{align}
\noindent This is the corrected (regularized) entropy due to the introduction of brick-wall. In this expression, the logarithmic correction is followed by the inverse area corrections of the black hole entropy. {An important observation in the above expression is that the subleading order logarithmic correction in the black hole entropy comes with a positive sign. However, there is some debate regarding the sign, as most of the string theoretic and LQG corrections consider the subtraction of the logarithmic correction from the semiclassical Bekenstein-Hawking entropy \cite{Sen:2012dw}. But, it is true that some models in string theory consider Bekenstein-Hawking entropy an undercounting of the entropy and take the logarithmic correction as an addition. The interesting development in this direction is that of Barrow entropy \cite{Barrow:2020tzx}, where, motivated by the Covid-19 virus structure, considered that quantum-level fractal correction in the horizon of the black hole gives greater black hole entropy than the Bekenstein-Hawking entropy. The positive coefficient in the logarithmic  term perfectly aligns with the Barrow entropy as it gives the black hole entropy greater than the Bekenstein Hawking entropy which we are getting in the above result. }
\section{Brick Wall model for Reissner-Nordstr\"om black hole }\label{RN}
\noindent While the Schwarzschild solution describes a static spherically symmetric metric, the Reissner--Nordstr\"om metric describes a solution to the Einstein equation for  a static, spherically symmetric, and charged black hole, with the lapse function 
\begin{equation}
    f(r)=\left(1-\frac{2M}{r}+\frac{Q^2}{r^2}\right)~.
\end{equation}
The introduction of charge significantly changes the spacetime as unlike to the Schwarzschild case, Reissner-Nordstr\"om admits two horizon. The radius of this two horizon has the form,
\begin{equation}
    r_\pm=M\pm\sqrt{M^2-Q^2} \ .
\end{equation}
These relations can be used to replace the mass ($M$) and change ($Q$) in terms of the inner ($r_-$) and outer ($r_+$) horizon radii. In terms of the horizon radii the lapse function takes the form
\begin{equation}
    f(r)=1-\frac{r_++r_-}{r}+\frac{r_+r_-}{r^2}~.
\end{equation}
Similar to the eq.\eqref{surface gravity schwarzschild} we can find the the surface gravity for RN black hole on the outer surface as
\begin{equation}
    \kappa=\frac{r_+-r_-}{2 r_+^2}~.
\end{equation}
Now this can be used to find out the free energy of the black hole by using the definition of temperature from eq.\eqref{horizon temperature} and finally the free energy from eq.\eqref{free energy}. Now using eq.\eqref{entropy} we can write the entropy for the RN black hole.
\begin{equation}\label{RN_entropy}
    S_{BW}=\frac{8\pi^3}{45\beta_H^3} \int_{{r_+}+h}^\infty dr \frac{r^2}{f(r)^2}
\end{equation}
To evaluate the exact form of the black hole we need to find out the integral in the above expression. For this we define $I\equiv \int_{{r_+}+h}^\infty dr \frac{r^2}{f(r)^2}$, which reads 
\begin{widetext}
\begin{align}\label{I_1}
    I&=\int_{{r_+}+h}^\infty dr \frac{r^2}{f(r)^2}
    =\int_h^\infty dx \frac{(x+r_+)^6}{x^2(x+r_+-r_-)^2}\nonumber \\
    &=\Bigg[\frac{x^3}{3}+(2r_++r_-)x^2+3\paren{2r_+^2+2r_+r_-+r_-^2}x-\frac{r_+^6}{\paren{r_+-r_-}^2x}+\frac{2r_+^5\paren{2r_+-3r_-}}{\paren{r_+-r_-}^3}\ln x\nonumber \\&~~~ -\frac{r_-^6}{\paren{r_+-r_-}^2\paren{x+r_+-r_-}}-\frac{2r_-^5\paren{2r_--3r_+}}{\paren{r_+-r_-}^3}\ln \paren{x+r_+-r_-}\left.\Bigg]\right|_h^\infty~.
\end{align}
\end{widetext}
If we use this expression and substitute back to the equation \eqref{RN_entropy} and after removing the divergent part that comes due to $L \rightarrow \infty $, we obtain the regularized entropy in terms of the brickwall width $h$
\begin{widetext}
\begin{align}\label{RNentropy}
    S_{BW}=&\frac{(1-\frac{r_-}{r_+})^3}{360 r_+^3}\biggr[\frac{r_+^4}{(1-\frac{r_-}{r_+})^2 h}-3(2r_+^2+2r_+r_-+r_-^2)h-2r_+(1+\frac{r_-}{r_+})h^2-\frac{h^3}{3}-\frac{4r_+^3 (1-\frac{3r_-}{2r_+})}{(1-\frac{r_-}{r_+})^3}\ln{h}\nonumber \\ &+\frac{r_-^6}{r_+^2(1-\frac{r_-}{r_+})^2(h+r_+-r_-)}-\frac{6r_-^5(1-\frac{2r_-}{3r_+})}{r_+^2(1-\frac{r_-}{r_+})^3}\ln{(h+r_+-r_-)} \biggr]~.
\end{align}
\end{widetext}
Now to get the form of the Brick wall width $h$, we equate the leading order term with the Bekenstein-Hawking entropy, mathematically,
\begin{align}
        &\frac{r_+ \paren{1-\frac{r_-}{r_+}}}{360 h}=\frac{\pi r_+^2}{l_p^2}\nonumber \\
        \implies &h=\frac{\paren{1-\frac{r_-}{r_+}}l_p^2}{360\pi r_+}~.
\end{align}
{In terms of the horizon area, it can be expressed as
\begin{align}
    h=\frac{l_P^2 \left(1-\sqrt{\frac{A_-}{A_+}}\right)}{180\sqrt{\pi A_+}}~.
\end{align}}
One can easily check that this form of the brick wall width $h$ smoothly matches with the Schwarzschild one for  $Q\rightarrow0$ that is, $r_-\rightarrow0$. Now using the above form of $h$ and inserting this back to the eq.\eqref{RNentropy} we got the sub leading order corrections in the entropy, which reads

\begin{widetext}
   {
    \begin{align}
       S_{BW} \simeq  &\frac{A_+}{4l_p^2}+\frac{1}{180}\Bigg(1-\frac{3}{2}\sqrt{\frac{A_-}{A_+}}\Bigg)\ln{\left(\frac{A_+}{4l_p^2}\right)} - \frac{l_P^2 \biggr(1-\sqrt{\frac{A_-}{A_+}} \biggr)^4}{(30\times 180)A_+}- \frac{l_P^4 \biggr(1-\sqrt{\frac{A_-}{A_+}} \biggr)^6}{(45\times (180)^2)A_+^2}-\frac{13}{1080}\Bigg(1-3\sqrt{\frac{A_-}{A_+}}\Bigg)\nonumber \\&-\frac{1}{90}\Bigg(1-\frac{3}{2}\sqrt{\frac{A_-}{A_+}}\Bigg)\ln{\Bigg(\frac{l_P}{360\sqrt{\pi}}\Big(1-\sqrt{\frac{A_-}{A_+}}\Big)\Bigg)}
    \end{align}}
\end{widetext}
Here, $A_{+}$ and $A_{-}$ are the areas corresponding to $r_{+}$ and $r_{-}$. The above form of entropy is interesting, as one can see that it has logarithmic terms. In the leading order the coefficient of the logarithmic correction is $\frac{1}{180}$. Logarithmic corrections in black hole entropy originate from other theories, such as loop quantum gravity, string theory, Euclidean quantum gravity, and generalized uncertainty principle theory. From the above analysis, we were able to produce the same type of logarithmic correction in the sub-leading order using the brick wall model.     

\section{Brick Wall model for Quantum corrected black holes }\label{RG improved}
\noindent In this section, we shall discuss the brick wall model in the renormalization group improved black hole geometry.\\
\noindent The general theory of relativity breaks down at a very small length scale due to its non-renormalizability, and hence the theory needed some modifications. The theoretical way of embodying nonperturbative quantum corrections to the solution of Einstein field equations is by considering the exact renormalization group flow equation (RGFE). This idea is based on the fact that the general theory of relativity is a theory at low energies coming from the theory which is valid at very high energy scales. The quantum improvement of Einstein's gravity is generally done in three different ways \cite{Reuter,Reuter2,ReuterTurian,
BonanoReuter,BonanoReuter2,Percacci}. The first way is to substitute running couplings directly into the metric. The second way is to improve the field equations and the third way is to improve the theory in the action level. The easiest way to incorporate this is to substitute the quantum corrections directly into the metric components. The motivation for doing so has been discussed in \cite{Ishibashi,jsg1,jsg2}. The construction of quantum improved geometry is carried out according to the following prescriptions \cite{BonanoReuter, StockPawlowski, Platania2, ReuterWeyer}:
\begin{enumerate}
\item From the exact RGFE, obtain the scale dependent coupling constants.
\item Express the UV cut-off scale as a function of radial coordinates in spherically symmetric geometries. This scale identification can be done in two ways:
\begin{enumerate}
    \item Based on scalars constructed out of curvature entities like the Ricci scalar, $R_{\alpha\beta}R^{\alpha\beta}$, or the Kretschmann scalar, $\mathcal{K}=R_{\alpha\beta\gamma\delta}R^{\alpha\beta\gamma\delta}$.
    \item Based on the UV fixed point that separates a weak-coupling regime from a strong-coupling regime.
\end{enumerate}
\item Include the quantum improvements which also can be done in three ways:
\begin{enumerate}
    \item The first way is to substitute the running couplings directly into the components of metric tensor.
    \item The second way is to impose the quantum corrections into the field equations. To obtain the quantum improved Einstein field equations, one can put the running gravitational constant and cosmological constant as
    \begin{equation}
    G_{\mu \nu} = 8\pi G(x) T_{\mu \nu}-\Lambda(x)g_{\mu \nu}~.
\end{equation}
The flow of the matter coupling has been ignored here.
    \item The third way is to improve the geometry in the action level. The quantum improved Einstein-Hilbert action has the form as
    \begin{equation}
    \mathcal{S}_{EH}=\frac{1}{16\pi}\int \frac{d^{4}x}{G(x)}\left(R-2\Lambda(x) \right)~.
\end{equation}
\end{enumerate}
The simplest way to do that is to improve it in the solution level directly. \\
\noindent In 2000, this approach was first applied on Schwarzschild geometry \cite{BonanoReuter} where the usual Newton's gravitational constant $G_{0}$ was replaced by the flow of the gravitational constant $G(r)$. Using the functional renormalization group in $U(1)$-coupled gravity theories, the $U(1)$ gauge completion has been done in \cite{EichhornVersteeg}. Using this method, the renormalization group improved Reissner-Nordstr\"{o}m black hole can be obtained, where the Newton's gravitational constant and the charge of the black hole are considered as running couplings in terms of momentum cut-off scale. The beta functions corresponding to these running couplings can be written as \cite{HarstReuter,EichhornVersteeg}
\begin{align}
k\frac{d\tilde{G}}{dk}&=2\tilde{G}\left(1-\frac{\tilde{G}}{4\pi\tilde{\alpha}}\right)\label{G beta func}\\
k\frac{de}{dk}&=\frac{e}{4\pi}\left(\frac{be^2}{4\pi^2}-\tilde{G}\right)\label{e beta func}
\end{align}
with $\tilde{G}(k)=k^2 G(k)$, $\tilde{\alpha}$ and $\beta$ are the parameters which specify the fixed points $\tilde{G}_*=4\pi\tilde{\alpha}$ and $e_*^2=(4\pi)^2\frac{\tilde{\alpha}}{b}$. In our work, we have considered a fixed charge, along with the flowing gravitational constant and in this case, $\tilde{\alpha}$ can be identified with the ``\textit{quantum gravity}" parameter $\tilde{\omega}'$ as \cite{Ishibashi,jsg2,RuizTurian}
\begin{equation}\label{alpha rel}
\tilde{\omega}'=\frac{1}{4\pi\tilde{\alpha}}.
\end{equation}
\noindent Now, solving eq.(\ref{G beta func}) and using eq.(\ref{alpha rel}), we get the flow of $G$ in terms of momentum cut-off scale as
\begin{equation}\label{flow of G}
G(k)=\frac{G_0}{1+\tilde{\omega}'k^2G_0}
\end{equation}
\noindent with $G_0 = G(k=0)$. In the near-horizon area ($r\gg 0$), the cut-off parameter $k$ has the form in terms of radial distance $r$ as \cite{EichhornVersteeg}
\begin{equation}\label{k as r}
k\simeq\frac{\xi}{r}.
\end{equation}
\noindent Now introducing $\tilde{\omega}=\xi^2\tilde{\omega}'$, we get the running gravitational constant as
\begin{equation}\label{running G}
G(r)=\frac{G_0}{1+\frac{\tilde{\omega}G_0}{r^2}}
\end{equation}
\noindent This is the flow of Newton's gravitational constant in terms of radial distance $r$.
\end{enumerate}
\subsection{Quantum corrected Schwarzschild black hole}
\noindent In this subsection, we shall calculate the regularized entropy for quantum corrected Schwarzschild black hole, for which the lapse function reads 
\begin{equation}
    f(r)=1-\frac{2G(r)M}{r} \ ,
\end{equation}
where $G(r)$ satisfies eq.(\ref{running G}) and hence, the lapse function takes the form, 
\begin{equation}
    f(r)=1-\frac{2MrG_0}{r^2+\tilde \omega G_0}
\end{equation}
To find the horizon, we shall use $f(r_H)=0$ and the horizon radius has the form as
\begin{equation}
    r_\pm=MG_0\pm\sqrt{M^2G_0^2-\tilde \omega G_0}
\end{equation}
\noindent Now, the surface temperature of the black hole is 
\begin{equation}
    T_H=\beta_H^{-1}=\frac{1}{4\pi}f'(r_H)=\frac{1}{4\pi}\frac{r_+-r_-}{r_+^2+\tilde\omega G_0}
\end{equation}
\noindent Now, we obtain the free energy for the quantum corrected Schwarzschild black hole using eq.(\ref{free energy}) as
\begin{align}
    F=-\frac{2\pi^3}{45\beta^4}\int_h^\infty dx \frac{\paren{x+r_+}^2\second{\paren{x+r_+}^2+\tilde\omega G_0}^2}{\second{\paren{x+r_+}^2+\tilde\omega G_0-2MG_0\paren{x+r_+}}^2}
\end{align}
Hence, using the definition in eq.(\ref{entropy}), we obtain the brick wall entropy as
\begin{equation}\label{Quantum_Schwarzschild_entropy}
    S_{BW}=\frac{8\pi^3}{45\beta_H^3} \int_h^\infty dx \frac{\paren{x+r_+}^2\second{\paren{x+r_+}^2+\tilde\omega G_0}^2}{\second{\paren{x+r_+}^2+\tilde\omega G_0-2MG_0\paren{x+r_+}}^2}~.
\end{equation}
So we have to find out the integration in the above equation, defining this as $I$, we have
    \begin{align}
        I&=\int_h^\infty dx \frac{\paren{x+r_+}^2\second{\paren{x+r_+}^2+\tilde\omega G_0}^2}{\second{\paren{x+r_+}^2+\tilde\omega G_0-2MG_0\paren{x+r_+}}^2}\nonumber \\& =\int_h^\infty dx \frac{\paren{x+r_+}^6\second{1+\frac{\tilde\omega G_0}{\paren{x+r_+}^2}}^2}{x^2\paren{x+r_+-r_-}^2}
    \end{align}
Now as $\forall x\in[h,\infty], \; \frac{\tilde\omega}{\paren{x+r_+}^2}\ll1 $ we can write 
\begin{align}\label{I}
    I &\simeq \int_h^\infty dx \frac{\paren{x+r_+}^6\second{1+\frac{2\tilde\omega G_0}{\paren{x+r_+}^2}}}{x^2\paren{x+r_+-r_-}^2} \nonumber \\
    &=\int_h^\infty dx\frac{\paren{x+r_+}^6}{x^2\paren{x+r_+-r_-}^2}+2\tilde\omega G_0\int_h^\infty dx\frac{\paren{x+r_+}^4}{x^2\paren{x+r_+-r_-}^2} \nonumber \\
    & \equiv I_1+I_2~.
\end{align} 
Here, $I_1$ has already been calculated in Eq. \eqref{I_1}, we have to calculate only the $I_2$ integration. Which has been done in the following.
\begin{widetext}
    \begin{align}\label{I_2}
        I_2&=\int_h^\infty dx \frac{\paren{x+r_+}^4}{x^2\paren{x+r_+-r_-}^2} \nonumber \\
        &=\third{-h-\frac{2r_+^3\paren{r_+-2r_-}}{\paren{r_+-r_-}^3}\ln{h}+\frac{r_+^4}{\paren{r_+-r_-}^2h}-\frac{2r_-^3\paren{2r_+-r_-}}{\paren{r_+-r_-}^3}\ln{\paren{h+r_+-r_-}}+\frac{r_-^4}{\paren{r_+-r_-}^2\paren{h+r_++r_-}}}
    \end{align}
\end{widetext}
Using eq.\eqref{I_1}and eq.\eqref{I_2} and inserting this in the eq.\eqref{Quantum_Schwarzschild_entropy} we can get the form of the entropy in term of brick wall width. Equaling the leading order term with the Bekenstein-Hawking entropy, we have
\begin{equation}
    \frac{8\pi^3}{45\beta_H^3}\third{\frac{r_+^4}{h\paren{1-\frac{r_-}{r_+}}^2}+\frac{2\tilde \omega G_0 r_+^2}{h\paren{1-\frac{r_-}{r_+}}^2}}=\frac{\pi r_+^2}{l_p^2}
\end{equation}
It is now straight forward to find out the form of the height of the brick wall $h$ upto $\mathcal{O}(\tilde{\omega})$,
\begin{equation}
    h\simeq \frac{l_p^2}{360\pi r_+}\paren{1-\frac{r_-}{r_+}}\paren{1-\frac{\tilde\omega G_0}{r_+^2}}~.
\end{equation}
In terms of the area which can be written as
\begin{equation}\label{h Quantum Schwarzschild}
    h\simeq\frac{l_p^2}{180\sqrt{\pi A_+}}\left(1-\sqrt{\frac{A_-}{A_+}}\right)\left(1-\frac{4\pi\tilde \omega G_0}{A_+}\right)
\end{equation}
Substituting the $h$, the form of the entropy of the black hole reads
\begin{widetext}
\begin{align}\label{entropy quantum corrected}
    S_{BW}&=\frac{\pi r_+^2}{l_p^2}+\frac{1}{360}\frac{(1-\frac{r_-}{r_+})^3}{r_+^3\paren{1+\frac{\tilde\omega G_0}{r_+^2}}^3}\Bigg[\frac{-h^3}{3}-(2r_++r_-)h^2-3\paren{2r_+^2+2r_+r_-+r_-^2}h \nonumber \\&-\frac{2r_+^5\paren{2r_+-3r_-}}{\paren{r_+-r_-}^3}\ln h+\frac{r_-^6}{\paren{r_+-r_-}^2\paren{h+r_+-r_-}}+\frac{2r_-^5\paren{2r_--3r_+}}{\paren{r_+-r_-}^3}\ln \paren{h+r_+-r_-}\nonumber \\&+2\tilde\omega G_0\Bigg\{-h-\frac{2r_+^3\paren{r_+-2r_-}}{\paren{r_+-r_-}^3}\ln{h}-\frac{2r_-^3\paren{2r_+-r_-}}{\paren{r_+-r_-}^3}\ln{\paren{h+r_+-r_-}}+\frac{r_-^4}{\paren{r_+-r_-}^2\paren{h+r_++r_-}}\Bigg\}\Bigg]
    \end{align}
{$\implies$
    \begin{align}
      S_{BW}\simeq &\frac{A_+}{4l_P^2}+\frac{1}{180}\biggr(1-\frac{8\pi \tilde{\omega}G_0}{A_+}\biggr)\ln{\biggr(\frac{A_+}{4l_P^2}\biggr)}+ \frac{1}{A_+}\biggr( -\frac{l_P^2 \biggr(1-\sqrt{\frac{A_-}{A_+}}\biggr)^4}{(30 \times 180)} +2\pi\tilde{\omega}G_0 \biggr) \nonumber \\ &+ \frac{l_P^2 \biggr(1-\sqrt{\frac{A_-}{A_+}}\biggr)^4}{45 A_+^2} \biggr(    \frac{2\pi\tilde{\omega}G_0 }{15}-\frac{l_P^2}{(180)^2} \biggr)-\frac{1}{90}\biggr(1-\frac{8\pi \tilde{\omega}G_0}{A_+}\biggr)\ln{\biggr(\frac{l_P}{360\sqrt{\pi}}\biggr(1-\sqrt{\frac{A_-}{A_+}}\biggr) \biggr)}  ~.
\end{align}}
where $h$ is derived in eq.\eqref{h Quantum Schwarzschild} and $r_\pm=\sqrt{\frac{A_\pm}{4\pi}}$
\end{widetext}
In the above expression of the entropy we can see that the first term is the Bekenstein-Hawking entropy and it corrected by other terms which are in the sub-leading order, logarithmic correction appears among them. In the leading order the coefficient of that logarithmic term is $1/4$ similar to the RN black hole without quantum correction. 
\subsection{ Quantum corrected Reissner--Nordstr\"om black hole }
\noindent We now proceed to apply the brick wall formalism in the context of RN black hole with the flow of the $G(r)$ considering the black hole has a charge $e$. Usually both $G$ and $e$ flow with the energy; however, in \cite{RuizTurian} it was shown that it is sufficient to work with the running coupling of Newton's constant, keeping the charge $e$ constant. For such case the lapse function reads
\begin{equation}
    f(r)=1-\frac{2G(r)M}{r}+\frac{G(r) e^2}{r^2}\ ,
\end{equation}
where $G(r)$ is given by eq.\eqref{running G}. So the lapse function reads 
\begin{equation}
    f(r)=1-\frac{2MrG_0}{r^2+\tilde\omega G_0}+\frac{e^2 G_0}{r^2+\tilde\omega G_0}
\end{equation} 
\begin{equation}
    r_\pm =MG_0\pm\sqrt{M^2 G_0^2-\paren{e^2+\tilde\omega}G_0}
\end{equation}
\begin{equation}
    f'(r)=\frac{r_+-r_-}{r_+^2+\tilde\omega}
\end{equation}
\begin{equation}
    T_H=\beta_H^{-1}=\frac{1}{4\pi}\frac{r_+-r_-}{r_+^2+\tilde\omega}
\end{equation}
Now we have to find out the integration $\int_{r_++h}^\infty dr\frac{r^2}{f^2(r)}$ to proceed further. 
Which can be done for this case as follows
\begin{align}
  &\int_{r_++h}^\infty dr\frac{r^2}{f^2(r)}= \int_{r_++h}^\infty dr\frac{r^2}{\left(1-\frac{2Mr}{r^2+\tilde \omega}+\frac{e^2}{r^2+\tilde \omega}\right)^2}\nonumber \\
&\overset{x\equiv r-r_+}{=}\int_h^\infty dx\frac{\left(x+r_+\right)^2\left(\left(x+r_+\right)^2+\tilde\omega\right)^2}{\paren{\left(x+r_+\right)^2+\tilde \omega-2M(x+r_+)+e^2}^2} \nonumber \\
&= \int_h^\infty dx \frac{\left(x+r_+\right)^6\paren{1+\frac{2\tilde\omega}{\left(x+r_+\right)^2}}}{x^2\paren{x+r_+-r_-}^2}\nonumber \\
&\simeq \int_h^\infty dx\frac{\paren{x+r_+}^6}{x^2\paren{x+r_+-r_-}^2}+2\tilde\omega\int_h^\infty dx\frac{\paren{x+r_+}^4}{x^2\paren{x+r_+-r_-}^2} 
\end{align}
The above form is same as of the eq.\eqref{I}. Hence the form of the entropy is same as of the quantum corrected Schwarzschild black hole that is eq.\eqref{entropy quantum corrected}. However the changes occurs in the form of the $r_\pm$, which incorporates the effect of the charge in the overall dynamics without changing the core form of the entropy.

\section{conclusion}\label{conclusion}
 In this paper, we have computed the quantum corrections to black hole entropy using the brick wall formalism. Specifically, by bypassing the near-horizon approximation of the lapse function, we have derived the non-approximated and robust expressions for the higher-order corrections to the black hole entropy. 
In this work, we have carried out the analysis for spherically symmetric black holes 
in 3+1 dimensions within a semiclassical framework. Here, gravity has been treated as a classical background, whereas other degrees of freedom inside the event horizon of the black hole are treated quantum mechanically. The density of the quantum states diverges near the event horizon due to the gravitational redshift, leading to the divergence of black hole entropy. To resolve this divergence, we have implemented 't Hooft's brick wall regularization formalism \cite{brickwall}, where a spatial boundary cutoff $h$ is introduced just outside the black hole event horizon at $r=r_h +h$. Utilizing the semiclassical WKB approximation within this boundary, the total free energy of the field was formulated directly as a function of the lapse function (eq.(\ref{free energy})).  
In section \ref{schwarzschild}, we performed this integral for the Schwarzschild metric, without taking the near horizon approximation of it, and then computed the canonical entropy in terms of the cutoff height $h$. Equating the leading order canonical entropy with the Bekenstein-Hawking entropy $S_{BH}$, we have estimated the height of the brickwall $h$. Substituting this value of $h$ into the expression of the canonical entropy $S_{BW}$, we find that the leading and subleading corrections correspond to the logarithmic and the inverse area corrections to the Bekenstein-Hawking entropy. In the later parts of this manuscript, we have repeated the analysis for different types of spherically symmetric black hole geometries. 
In section {\ref{RG improved}}, we have applied the brick wall approach to the quantum corrected geometries, where the Newton's gravitational constant flows with certain momentum cutoff scale. In these cases, the height of the brick wall we obtained is dependent on the quantum gravity parameter $\tilde{\omega}$. In this entire study, we have considered the  black holes to be non-extremal and applied the brick wall approach and found the entropy to have universal correction terms, along with finite coefficients. {Our analysis reveals a key feature of the quantum correction to the black hole entropy, namely, a positive coefficient in front of the subleading logarithmic term. While the sign of this coefficient remains a subject of controversy within the string theoretic and LQG studies, a positive correction aligns with the fractal horizon model introduced by Barrow in \cite{Barrow:2020tzx}, where the quantum gravitational effects increase the black hole entropy relative to the standard Bekenstein-Hawking entropy. In our analysis, the positive coefficient in the quantum corrected entropy aligns directly with the framework of Barrow entropy. }


\begin{thebibliography}{80}
\bibitem{Bekenstein1}
J. D. Bekenstein, ``\textit{Black holes and the second law}", \href{https://doi.org/10.1007/BF02757029}{Lett. Nuovo Cimento 4, 737–740 (1972)}.
\bibitem{Bekenstein2}
J. D. Bekenstein, ``\textit{Black Holes and Entropy}", \href{https://doi.org/10.1103/PhysRevD.7.2333}{Phys. Rev. D 7, 2333 (1973)}.
\bibitem{Hawking1}
S. W. Hawking, ``\textit{Black hole explosions?}", \href{https://doi.org/10.1038/248030a0}{Nature 248, 30 (1974)}.
\bibitem{Hawking2}
S. W. Hawking, ``\textit{Partile creation by black holes}", \href{https://doi.org/10.1007/BF02345020}{Commun. Math. Phys. 43, 199 (1975)}.
\bibitem{Hawking3}
S. W. Hawking, ``\textit{Black holes and thermodynamics}", \href{https://link.aps.org/doi/10.1103/PhysRevD.13.191}{Phys. Rev. D 13, 191 (1976)}.
\bibitem{strominger}
A. Strominger and C. Vafa, ``\textit{Microscopic origin of the Bekenstein-Hawking entropy}", \href{https://doi.org/10.1016/0370-2693(96)00345-0}{Phys. Lett. B 379, 99 (1996)}.
\bibitem{johnsonkhuri}
 C.V. Johnson, P.R. Khuri and R.C. Meyers, ``\textit{Entropy of 4D extremal black holes}", \href{https://doi.org/10.1016/0370-2693(96)00383-8}{Phys. Lett. B 378, 78
(1996)}.
\bibitem{maldacena}
J.M. Maldacena and A. Strominger, ``\textit{Statistical Entropy of Four-Dimensional Extremal Black Holes}", \href{https://doi.org/10.1103/PhysRevLett.77.428}{Phys. Rev. Lett. 77, 428 (1996)}.
\bibitem{banerjeegupta1}
 S. Banerjee, R.K. Gupta and A. Sen, ``\textit{Logarithmic corrections to extremal black hole entropy from quantum entropy function}", \href{https://doi.org/10.1007/JHEP03(2011)147}{J. High Energ. Phys. 2011, 147 (2011)}.
\bibitem{banerjeegupta2}
 S. Banerjee, R.K. Gupta, I. Mandal and A. Sen, ``\textit{Logarithmic corrections to N=4
 and N=8 black hole entropy: a one loop test of quantum gravity.}", \href{https://doi.org/10.1007/JHEP11(2011)143}{J. High Energ. Phys. 2011, 143 (2011)}.
\bibitem{callan}
 C.G. Callan and J.M. Maldacena, ``\textit{D-brane approach to black hole quantum mechanics}", \href{https://doi.org/10.1016/0550-3213(96)00225-8}{Nucl. Phys. B 472, 591 (1996)}.
 \bibitem{horowitz}
  G.T. Horowitz and A. Strominger, ``\textit{Counting States of Near-Extremal Black Holes}", \href{https://doi.org/10.1103/PhysRevLett.77.2368}{Phys. Rev. Lett. 77, 2368 (1996)}.
\bibitem{gibbons}
G.W. Gibbons and S.W. Hawking, ``\textit{Action integrals and partition functions in quantum gravity}", \href{https://doi.org/10.1103/PhysRevD.15.2752}{Phys. Rev. D 15, 2752 (1977)}.
\bibitem{york}
J.W. York, ``\textit{Black-hole thermodynamics and the Euclidean Einstein action}", \href{https://doi.org/10.1103/PhysRevD.33.2092}{Phys. Rev. D 33, 2092 (1986)}.
\bibitem{braden}
H.W. Braden, J.D. Brown, B.F. Whiting and J.W. York, ``\textit{Charged black hole in a grand canonical ensemble}", \href{https://doi.org/10.1103/PhysRevD.42.3376}{Phys. Rev. D 42, 3376 (1990)}.
\bibitem{alwis1}
 S.P. de Alwis and N. Ohta, ``\textit{On the Entropy of Quantum Fields in Black Hole Backgrounds}", \href{https://doi.org/10.48550/arXiv.hep-th/9412027}{arXiv:hep-th/9412027}.
 \bibitem{alwis2}
S.P. de Alwis and N. Ohta, ``\textit{Thermodynamics of quantum fields in black hole backgrounds}", \href{https://doi.org/10.1103/PhysRevD.52.3529}{Phys. Rev. D 52, 3529 (1995)}. 
\bibitem{sen}
A. Sen, ``\textit{Logarithmic corrections to Schwarzschild and other non-extremal black hole entropy in different dimensions}", \href{https://doi.org/10.1007/JHEP04(2013)156}{J. High Energ. Phys. 2013, 156 (2013)}.
\bibitem{carlip1}
S. Carlip, ``\textit{Near-Horizon Conformal Symmetry and Black Hole Entropy}", \href{https://doi.org/10.1103/PhysRevLett.88.241301}{Phys. Rev. Lett. 88, 241301 (2002)}.
\bibitem{carlip2}
S. Carlip, ``\textit{Logarithmic corrections to black hole entropy, from the Cardy formula}", \href{https://doi.org/10.1088/0264-9381/17/20/302}{Class. Quant. Grav. 17, 4175 (2000)}.
\bibitem{caicao}
R.-G. Cai, L.-M. Cao and N. Ohta, ``\textit{Black holes in gravity with conformal anomaly and logarithmic term in black hole entropy}", \href{https://doi.org/10.1007/JHEP04(2010)082}{J. High Energ. Phys. 2010, 82 (2010)}. 
\bibitem{srinivasan}
K. Srinivasan and T. Padmanabhan, ``\textit{Particle production and complex path analysis}", \href{https://doi.org/10.1103/PhysRevD.60.024007}{Phys. Rev. D 60, 024007 (1999)}.
\bibitem{banerjeemajhi}
 R. Banerjee and B.R. Majhi, ``\textit{Quantum tunneling beyond semiclassical approximation}", \href{https://doi.org/10.1088/1126-6708/2008/06/095}{J. High Energ. Phys. 2008, 95 (2008)}.
\bibitem{brickwall}
 G. 't Hooft, ``\textit{On the quantum structure of a black hole}", \href{https://doi.org/10.1016/0550-3213(85)90418-3}{Nucl. Phys. B 256, 727-745 (1985)}.
 \bibitem{susskind}
 L. Susskind and J. Uglum, ``\textit{Black hole entropy in canonical quantum gravity and superstring theory}", \href{https://doi.org/10.1103/PhysRevD.50.2700}{Phys. Rev. D 50, 2700 (1994)}.
 \bibitem{mann}
 R. B. Mann, L. Tarasov and A. Zelnikov, ``\textit{Brick walls for black holes}", \href{https://doi.org/10.1088/0264-9381/9/6/006}{Class. Quant. Grav. 9, 1487 (1992)}.
 \bibitem{demers}
 J.- G. Demers, R. Lafrance and R.C. Myers, ``\textit{Black hole entropy without brick walls}", \href{https://doi.org/10.1103/PhysRevD.52.2245}{Phys. Rev. D 52, 2245 (1995)}.
 \bibitem{barbon}
  J.L.F. Barbon and R. Emparan, ``\textit{Quantum black hole entropy and Newton constant renormalization}", \href{https://doi.org/10.1103/PhysRevD.52.4527}{Phys. Rev. D 52, 4527 (1995)}.
\bibitem{kimkulkarni}
W. Kim and S. Kulkarni, ``\textit{Higher order WKB corrections to black hole entropy in brick wall formalism}", \href{https://doi.org/10.1140/epjc/s10052-013-2398-6}{Eur. Phys. J. C 73, 2398 (2013)}.
\bibitem{winstanley}
E. Winstanley, ``\textit{Renormalized black hole entropy in anti–de Sitter space via the ‘‘brick wall’’ method}", \href{https://doi.org/10.1103/PhysRevD.63.084013}{Phys. Rev. D 63, 084013 (2001)}.
\bibitem{kimkim}
S. P. Kim, S. K. Kim, K.-S. Soh, and J. H. Yee, ``\textit{Remarks on Renormalization of Black Hole Entropy}", \href{https://doi.org/10.1142/S0217751X97002802}{Int. J. Mod.
Phys. A 12, 5223 (1997)}.
\bibitem{ghoshmitra1}
 A. Ghosh and P. Mitra, ``\textit{Entropy in Dilatonic Black Hole Background}", \href{https://doi.org/10.1103/PhysRevLett.73.2521}{Phys. Rev. Lett. 73, 2521 (1994)}.
\bibitem{caizhang}
 R.-G. Cai and Y.-Z. Zhang, ``\textit{Entropy of scalar fields in Reissner–Nordstr\``om–(Anti-)de sitter spacetimes}", \href{https://doi.org/10.1142/S0217732396002010}{Mod. Phys. Lett. A 11, 2027 (1996)}.
\bibitem{ghoshmitra2}
A. Ghosh and P. Mitra, ``\textit{Entropy for extremal Reissner-Nordstrom black holes}", \href{https://doi.org/10.1016/0370-2693(95)00922-8}{Phys. Lett. B 357, 295 (1995)}.


  
\bibitem{Kaul}
R. K. Kaul and P. Majumdar, ``\textit{Logarithmic Correction to the Bekenstein-Hawking Entropy}", \href{ https://doi.org/10.1103/PhysRevLett.84.5255}{Phys. Rev. Lett. 84, 5255 (2000)  }.
\bibitem{DasMajumdar}
S. Das, P. Majumdar and R. K. Bhaduri, ``\textit{General logarithmic corrections to black-hole entropy}", \href{https://doi.org/10.1088/0264-9381/19/9/302}{Class. Quantum Grav. 19, 2355 (2002) }
\bibitem{gup1}
S. Gangopadhyay, A. Dutta and A. Saha, ``\textit{Generalized uncertainty principle and black hole thermodynamics}", \href{https://doi.org/10.1007/s10714-013-1661-3}{Gen. Rel.Grav. 46, 1661 (2014) }.
\bibitem{gup2}
S. Gangopadhyay and A. Dutta, ``\textit{Black hole thermodynamics and generalized uncertainty principle with higher order terms in momentum uncertainty}", \href{https://doi.org/10.1155/2018/7450607}{Adv. High Energy Phys. 2018, 7450607 (2018) }.
\bibitem{gup3}
A. Dutta and S. Gangopadhyay, ``\textit{Thermodynamics of black holes and the symmetric generalized uncertainty principle}", \href{https://doi.org/10.1007/s10773-015-2907-5}{	Int .J . Theor. Phys. 55, 2746 (2016) }.
\bibitem{Sen:2012dw}
A.~Sen,
``\textit{Logarithmic Corrections to Schwarzschild and Other Non-extremal Black Hole Entropy in Different Dimensions}",\href{https://doi.org/10.1007/JHEP04(2013)156}{
JHEP \textbf{04} (2013), 156}.
\bibitem{Barrow:2020tzx}
J.~D.~Barrow,
\textit{``The Area of a Rough Black Hole'',}
\href{https://doi.org/10.1016/j.physletb.2020.135643}{Phys. Lett. B \textbf{808}, 135643 (2020)}.


 
 \bibitem{Reuter}
M. Reuter, ``\textit{Nonperturbative evolution equation for quantum gravity}", \href{https://doi.org/10.1103/PhysRevD.57.971}{Phys. Rev. D 57, 971 (1998)}.
\bibitem{Reuter2}
M. Reuter and F. Saueressig, ``\textit{Quantum Gravity and the Functional Renormalization Group; The Road towards Asymptotic Safety}", Cambridge University Press (2019). \href{https://doi.org/10.1017/9781316227596}{Online Link}.
\bibitem{Percacci}
R. Percacci, ``\textit{An Introduction to Covariant Quantum Gravity and Asymptotic Safety (2017)}", \href{https://doi.org/10.1142/10369 }{Online Link}.
\bibitem{BonanoReuter}
A. Bonanno and M. Reuter, ``\textit{Renormalization group improved black hole spacetimes}", \href{https://doi.org/10.1103/PhysRevD.62.043008}{Phys. Rev. D 62, 043008 (2000)}.
\bibitem{BonanoReuter2}
A. Bonanno and M. Reuter, ``\textit{Spacetime structure of an evaporating black hole in quantum gravity}", \href{https://doi.org/10.1103/PhysRevD.73.083005}{Phys. Rev. D 73, 083005 (2006)}.
\bibitem{ReuterTurian}
M. Reuter and E. Tuiran, ``\textit{Quantum gravity effects in the Kerr spacetime}", \href{https://doi.org/10.1103/PhysRevD.83.044041}{Phys. Rev. D 83, 044041 (2011)}.
\bibitem{Ishibashi}
A. Ishibashi, N. Ohta and D. Yamaguchi, ``\textit{Quantum improved charged black holes}", \href{https://doi.org/10.1103/PhysRevD.104.066016}{Phys. Rev. D 104, 066016 (2021) }.
\bibitem{jsg1}
A. Jana, S. Sen and S. Gangopadhyay, ``\textit{Atom falling into a quantum corrected charged black hole and HBAR entropy}", \href{https://doi.org/10.1103/PhysRevD.110.026029}{Phys. Rev. D 110, 026029 (2024) }.
\bibitem{jsg2}
A. Jana, S. Sen and S. Gangopadhyay, ``\textit{Inverse logarithmic correction in the horizon brightened acceleration radiation entropy of an atom falling into a renormalization group improved charged black hole}", \href{ https://doi.org/10.1103/PhysRevD.111.085017}{Phys. Rev. D 111, 085017 (2025)}.
\bibitem{StockPawlowski}
J. M. Pawlowski and D. Stock, ``\textit{Quantum-improved Schwarzschild-(A)dS and Kerr-(A)dS spacetimes}", \href{https://doi.org/10.1103/PhysRevD.98.106008}{Phys. Rev. D 98, 106008 (2018)}.
\bibitem{Platania2}
A. Platania, ``\textit{From Renormalization Group Flows to Cosmology}", \href{https://www.frontiersin.org/articles/10.3389/fphy.2020.00188/full}{Front. Phys. 8, 188 (2020)}.
\bibitem{ReuterWeyer}
M. Reuter and H. Weyer, ``\textit{Running Newton constant, improved gravitational actions, and galaxy rotation curves}", \href{https://link.aps.org/doi/10.1103/PhysRevD.70.124028}{Phys. Rev. D 70, 124028 (2004)}.
\bibitem{EichhornVersteeg}
A. Eichhorn and F. Versteegen, ``\textit{Upper bound on the Abelian gauge coupling from asymptotic safety}", \href{https://link.springer.com/article/10.1007/JHEP01(2018)030}{J. High Energy Phys. 01, 030 (2018)}.
\bibitem{HarstReuter}
U. Harst and M. Reuter, ``\textit{QED coupled to QEG}", \href{https://link.springer.com/article/10.1007/JHEP05(2011)119}{J. High Energy Phys. 05, 119 (2011)}.
\bibitem{RuizTurian}
O. Ruiz and E. Tuiran, ``\textit{Nonperturbative quantum correction to the Reissner-Nordstr\"{o}m spacetime with running Newton’s constant}", \href{https://link.aps.org/doi/10.1103/PhysRevD.107.066003}{Phys. Rev. D 107, 066003 (2023)}.
\end{thebibliography}
\end{document}